\documentclass[10pt,conference]{IEEEtran}
\IEEEoverridecommandlockouts
\usepackage{cite}
\usepackage{enumitem}
\usepackage{amsmath,amssymb,amsfonts}
\usepackage{algorithmic}
\usepackage{graphicx}
\usepackage{textcomp}
\usepackage{xcolor}
\usepackage[capitalise]{cleveref}
\usepackage{xcolor}

\usepackage{balance}
\begin{document}

\title{Multi-Agent SAC Enabled Beamforming Design for Joint Secret Key Generation and Data Transmission\\
\thanks{This research was supported in part by the Shandong Provincial Natural Science Foundation under Grant ZR2023LZH003, the Major Basic Research Program of Shandong Provincial Natural Science Foundation under Grant ZR2025ZD18, and the Joint Key Funds of National Natural Science Foundation of China under Grant U23A20302.}
\author{
\IEEEauthorblockN{Ziao Wang\IEEEauthorrefmark{1}, Zheng Dong\IEEEauthorrefmark{1}, He (Henry) Chen\IEEEauthorrefmark{2}, Jun Chen\IEEEauthorrefmark{3}, and Dongxiao Yu\IEEEauthorrefmark{4}}
    \IEEEauthorblockA{
    \IEEEauthorrefmark{1}School of Information Science and Engineering, Shandong University, Qingdao, 266237, China\\
    \IEEEauthorrefmark{2}Department of Information Engineering, the Chinese University of Hong Kong, Hong Kong SAR, 999077, China\\
    \IEEEauthorrefmark{3}Department of Electrical and Computer Engineering, McMaster University, Hamilton, Ontario, L8S 4K1, Canada\\
    \IEEEauthorrefmark{4}School of Computer Science and Technology, Shandong University, Qingdao, 266237, China\\
    Emails: \{zhengdong, dxyu\}@sdu.edu.cn, he.chen@ie.cuhk.edu.hk, chenjun@mcmaster.ca
         \vspace{-10pt} 
    }
}
}

\maketitle

\begin{abstract}
Physical layer key generation (PLKG) has emerged as a promising solution for achieving highly secured and low-latency key distribution, offering information‑theoretic security that is inherently resilient to quantum attacks. However, simultaneously ensuring a high data transmission rate and a high secret key generation rate  under eavesdropping attacks  remains a major challenge. 
In time-division duplex (TDD) systems with multiple antennas, we derive closed-form expressions for both rates by modeling the legitimate channel as a time-correlated  autoregressive (AR) process.
This formulation leads to a highly nonconvex and time-coupled optimization problem, rendering traditional optimization methods ineffective.
To address this issue, we propose a multi-agent soft actor–critic (SAC) framework equipped with a long short-term memory (LSTM) adversary prediction module to cope with the partial observability of the eavesdropper’s mode.
Simulation results demonstrate that the proposed approach achieves superior performance compared with other benchmark algorithms, while effectively balancing the trade-off between secret key generation rate and data transmission rate. The results also confirm the robustness of the proposed framework against intelligent eavesdropping and partial observation uncertainty.
\end{abstract}


\section{Introduction}
With the development of sixth-generation (6G) wireless networks, we are stepping into a new era of the Internet of Things (IoT) with a requirement on intrinsic security, making data security more critical than ever before~\cite{10807044}. Traditional encryption-based techniques for key distribution predominantly rely on assumptions of computational complexity, which have increasingly revealed vulnerabilities in the context of progressively powerful classical computing attacks. Once large-scale quantum computers become a reality, these cryptographic systems could be rapidly compromised. 

Against this backdrop, physical layer key generation (PLKG) has emerged as a promising alternative for key distribution in dynamic environments~\cite{zhang2024artificial}. By leveraging the inherent randomness and reciprocity of wireless channels, PLKG enables communicating pairs to generate secret keys from their shared channel state information (CSI)~\cite{11027013}. With its solid foundation in information-theoretic security and inherent resistance to quantum computing attacks, PLKG offers a forward-secure approach particularly appealing in the post-quantum era. As a result, PLKG is gradually being recognized as a viable and efficient solution to achieve highly secure and low-latency secret key distribution in future 6G systems~\cite{10379509}.

In practical multi-antenna systems, beamforming plays a critical role in enhancing communication reliability and spectral efficiency.
For secret key design, by concentrating signal energy toward legitimate receivers and minimizing leakage to potential eavesdroppers, beamforming not only enhances channel reciprocity but also reinforces the confidentiality of the secret key generation process~\cite{10869337}. 
In addition, it facilitates a more precise channel estimation, which is essential to generate highly correlated cryptographic secret keys between authorized communication parties. By optimizing the beam direction and adaptively adjusting the beam width, a balance can be achieved between improving the accuracy of channel estimation and increasing the beam scanning delay, thus effectively utilizing the spatial degrees-of-freedom (DoF) to improve the capacity and efficiency of secret key generation~\cite{10845874}.

Such PLKG and beamforming mechanisms are commonly implemented in multi-antenna time-division duplex (TDD) systems, which are widely adopted in modern wireless standards such as 4G, 5G, and future 6G networks due to their inherent channel reciprocity~\cite{9205612}. 
Nevertheless, due to the non-simultaneous channel probing at the two legitimate devices in TDD systems, channel observations are typically acquired with a non-negligible duplexing delay, which weakens instantaneous channel reciprocity and degrades the achievable secret key generation rate.
Autoregressive (AR) models provide an effective way to characterize channel memory by capturing temporal correlation across consecutive time slots~\cite{9427230}. 

However, the presence of channel temporal correlation further complicates the design, rendering conventional optimization approaches intractable or overly conservative.
In contrast, deep reinforcement learning (DRL) provides a model-free framework that enables agents to directly learn adaptive decision policies through online interaction with the environment, without requiring explicit knowledge on the channel distribution or a closed-form solution~\cite{10440494,10620776}.
Motivated by these challenges, learning-based control mechanisms based on DRL naturally emerge as a promising solution for joint beamforming design in simultaneous secret key generation and data transmission systems. Among DRL algorithms, soft actor–critic (SAC) is an efficient method that learns a stochastic policy and has demonstrated state-of-the-art performance across a wide range of standard benchmark environments~\cite{anbazhagan2024next}. Based on the maximum entropy framework, SAC effectively balances exploration and convergence stability by optimizing a stochastic policy with adaptive entropy regularization.

In this paper, we consider the beamforming design in a multi-antenna system aided by multi-agent SAC, which strikes a balance between secret key generation and data transmission, offering a robust and near-optimal solution.
Our main contributions are as follows:
\begin{itemize}
    \item  
    We proposed a multi-agent SAC framework to address the non-convexity of the beamforming optimization in joint secret key generation and data transmission under intelligent eavesdropping attacks.
    \item 
    To tackle the issue of partial observability regarding channel fading and eavesdroppers, we introduced a  long short-term memory (LSTM) adversary prediction module, significantly enhancing system robustness and overall performance. 
    \item 
    We modeled the CSI errors caused by duplex delay using an AR(1) channel model and derived closed-form expressions for the secret key generation rate that explicitly account for the resulting temporal correlation.
\end{itemize}
$\mathit{Notations}$: Boldface letters denote matrices, vectors, or sets, while regular letters represent scalar quantities. $\mathbb{C} ^{M\times N}$ denotes the set of $M \times N$ dimensional complex matrices or vectors. $[\cdot ]^T,[\cdot ]^*$ and $[\cdot ]^H$ denote the transposition, conjugation, and Hermitian operations, respectively. The modulus of a scalar $x$ is denoted as $|x|$, the norm of a vector $\boldsymbol{x}$ is represented by $\left \|\boldsymbol{x}\right \|$, while the determinant of matrix $\boldsymbol{X}$ is represented as $\mathrm{det}\left (\boldsymbol{X}\right )$. $\mathbb{E}\{\cdot\}$ and $\mathfrak{R}\{\cdot\}$ denote expectation and real-part operators, respectively.

\section{System Model}
\begin{figure}[t]
    \centering 
    \includegraphics[width=0.4\textwidth]{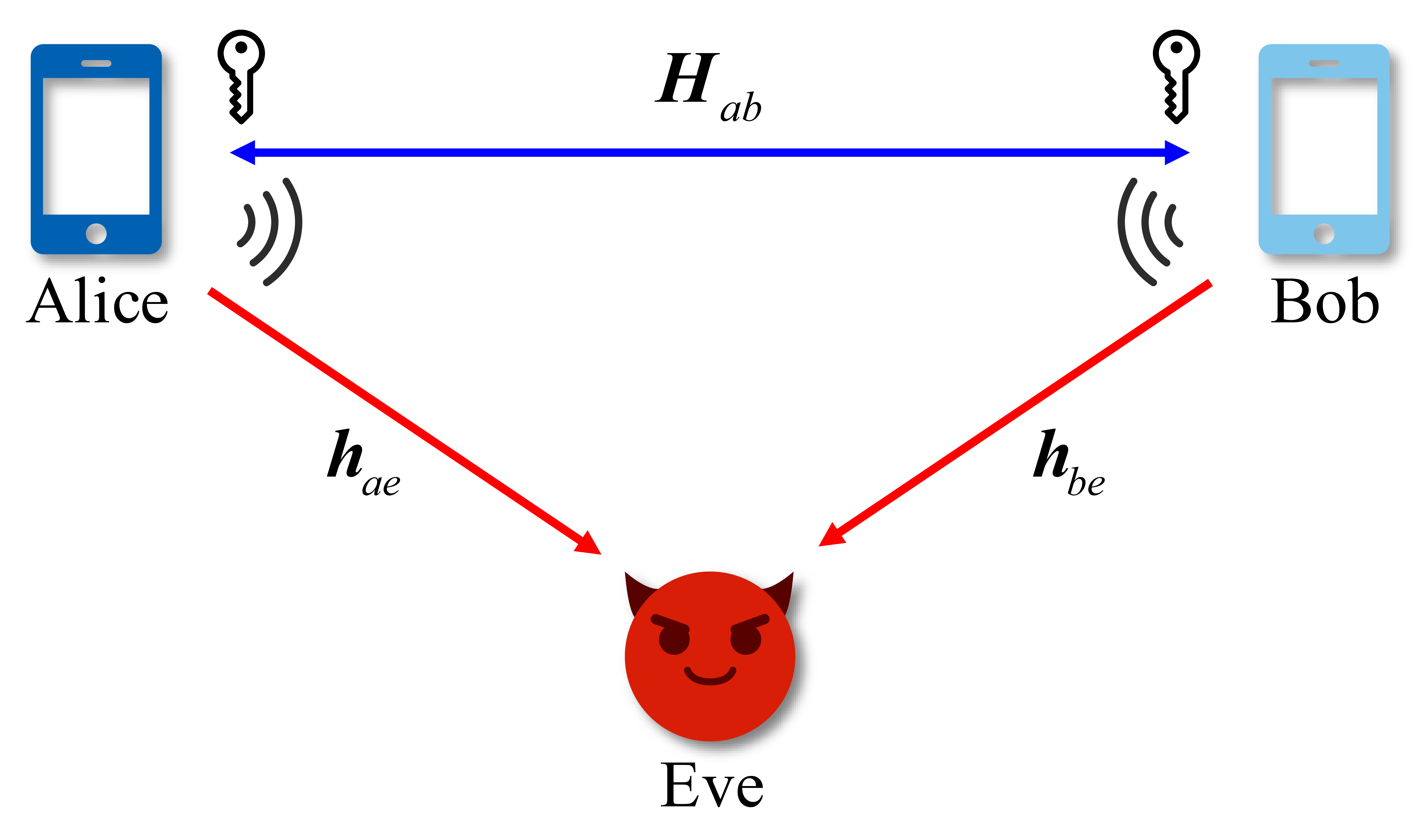} 
   \caption{The system model.} 
    \label{fig1} 
    \vspace{-10pt}
\end{figure}

We consider a multi-antenna TDD system for the bidirectional communications, as shown in~\cref{fig1}. 
In this system, legitimate
devices Alice and Bob communicate with each other, while passive node Eve attempts to eavesdrop the communication between Alice and Bob. We assume that both Alice and Bob have $N$ antennas, while Eve is
a single antenna device. Alice and Bob seek to not only transmit data but also generate physical-layer secret keys, all under the threat of Eve, who maliciously eavesdrops via the eavesdropping channel. The system employs a single data stream for secure transmission,  
where Alice and Bob utilize beamforming to enhance both communication and PLKG performance. The secret key generation is performed intermittently with intervals greater than the channel coherence time to ensure security.

\subsection{Channel Model}
The direct legitimate channel and eavesdropping channels are denoted by $\boldsymbol{H}_{ab} \in \mathbb{C} ^{N\times N}$  and $\boldsymbol{h}_{ae}, \boldsymbol{h}_{be} \in \mathbb{C} ^{N\times 1}$, respectively.
Beamforming is employed at both the Alice and Bob terminals to enhance communication reliability and PLKG performance.
The signal received by Bob can be denoted as
\begin{equation}
y_{b}=\sqrt{P_a}\boldsymbol{w}_{b}^{H}\boldsymbol{H}_{ab}\boldsymbol{w}_{a}{x}_{a}+\boldsymbol{w}_b^H \boldsymbol{z}_b=\tilde{h}_{ab}x_{a}+z_{b},
\end{equation}
where
$\boldsymbol{w}_{a}=[u_{1}e^{j\psi_1},u_{2}e^{j\psi_2},...,u_{N}e^{j\psi_N}]^T \in \mathbb{C} ^{N\times 1}$ and $\boldsymbol{w}_{b}=[v_{1}e^{j\phi_1},v_{2}e^{j\phi_2},...,v_{N}e^{j\phi_N}]^T \in \mathbb{C} ^{N\times 1}$ are the unit-norm beamforming vectors for Alice and Bob  with the amplitudes $u_{n},v_{n} \in \mathbb{R}^+$ and the phase shift $\psi_{n},\phi_{n} \in [0,2\pi)$, $n \in \{1,...,N\}$,
${x}_{a}$ is Alice's transmission signal,
$\boldsymbol{z}_b\in \mathbb{C} ^{N\times 1}$ is the additive noise vector whose entries are zero-mean complex Gaussian random variables with variance $\sigma_z^2$, $z_b =\boldsymbol{w}_b^H \boldsymbol{z}_b$ is the received noise,
$\tilde{h}_{ab}=\sqrt{P_a}\boldsymbol{w}_{b}^{H}\boldsymbol{H}_{ab}\boldsymbol{w}_{a}$ is the equivalent channel, $P_a$ is the transmit power of Alice. Under channel reciprocity, the received signal from Bob to Alice is
\begin{equation}
y_{a}=\sqrt{P_b}\boldsymbol{w}_{a}^{H}\boldsymbol{H}_{ba}\boldsymbol{w}_{b}{x}_{b}+\boldsymbol{w}_a^H \boldsymbol{z}_a=\tilde{h}_{ba}x_{b}+z_{a}.
\end{equation}
Likewise, the signals eavesdropped by Eve from Alice and Bob can be expressed as
\begin{subequations}
\begin{align}
&{y}_{ea}=\sqrt{P_a}\boldsymbol{h}_{ae}^{H}\boldsymbol{w}_{a}{x}_{a}+{z}_{e}=\tilde{h}_{ae}x_{a}+z_{e},\\
&{y}_{eb}=\sqrt{P_b}\boldsymbol{h}_{be}^{H}\boldsymbol{w}_{b}{x}_{b}+{z}_{e}=\tilde{h}_{be}x_{b}+z_{e}. 
\end{align}
\end{subequations}

We consider a narrowband flat-fading channel, where the channel gains $\boldsymbol{H}_{ab,i}, \boldsymbol{h}_{ae}$, and $\boldsymbol{h}_{be}$ are independent and identically distributed (i.i.d.) circularly-symmetric complex Gaussian random variables with zero mean and unit variance.
In practical TDD systems, the two reciprocal channel estimates for secret key generation are often not identical due to the time lag during the bidirectional channel estimation, leading to CSI mismatches. To characterize the temporal correlation of the channel, we employ an AR(1) model to describe the time-domain evolution of the legitimate channels:
\begin{equation}
\boldsymbol{H}_{ab}^t=\rho \boldsymbol{H}_{ab}^{t-1}+\boldsymbol{Z}_{ab}^t,
\end{equation}
where $\rho\in[0,1]$ is the autoregressive coefficient,  
and $\boldsymbol{Z}_{ab}^t$
denotes the matrix of complex Gaussian white noise innovations with mean zero and variance $\sigma_\zeta^2$, independent of $\boldsymbol{H}_{ab}^{t-1}$.
Meanwhile, the equivalent channel can be represented as
\begin{equation}
\tilde{h}_{ab}^{t}=
\rho\tilde{h}_{ab}^{t-1}+\tilde{z}_{ab}^t,
\end{equation}
where $\tilde{z}_{t}=\sqrt{P_a}\boldsymbol{w}_{b}^{H}\boldsymbol{Z}_{ab}^{t}\boldsymbol{w}_{a}$ denotes the equivalent innovation noise term. Hence, from a statistical perspective, the equivalent channel continues to follow an AR(1) process.

\subsection{Attack Model}
Although PLKG is inherently resistant to eavesdropping,  it is still necessary to consider the behavior of eavesdroppers to prevent the potential risk of exposure of relevant parameters in long-term statistical analysis. To model a more sophisticated threat, we consider Eve to be an intelligent attacker operating in two distinct modes: sleeping and eavesdropping. We refer to this as intelligent eavesdropping attacks and represent these behaviors as $\xi^t\in\{e,s\}$, where $e$ represents eavesdropping mode and $s$ represents sleeping mode.
To conserve power, Eve operates in sleeping mode when the wiretap channel quality is poor. Conversely, it switches to eavesdropping mode to monitor the PLKG process when the channel quality is favorable. We set the gain threshold of the eavesdropping channel as $\tau$, and the mode of Eve $\xi$ can be defined as
\begin{equation}
\xi^t =
\left\{
\begin{array}{@{}c@{\quad}c@{}}
e, & \text{if } \min\{||\boldsymbol{h}_{ae}^t||, ||\boldsymbol{h}_{be}^t||\} \ge \tau,\\[4pt]
s, & \text{otherwise}.
\end{array}
\right.
\end{equation}

\section{Problem Formulation}
In this section, we first derive the beamforming-assisted secret key generation and data transmission rates. Subsequently, we formulate the optimization problem to maximize the weighted sum of the rates.
\subsection{Secret Key Generation Rate}
The PLKG method typically incorporates CSI as an intrinsic complementary random source in the secret key generation process, which comprises channel probing, quantization, information reconciliation, and privacy amplification.
The secret key generation interval $T_{k}$ is set to exceed the channel coherence time $T_{c}$ to reduce correlation between channel samples.

We consider Alice as the active initiator for secret key generation, and analyze the secret key generation rate $R_{k}$ from the perspective of Bob. 
Based on Maurer's secret key capacity theory~\cite{maurer2000information}, the theoretical upper bound of $R_{k}$ is determined by the conditional mutual information $I(\tilde{h}_{ab};\tilde{h}_{ba})$.
Since PLKG relies on Alice and Bob's correlated
observations of the same random channel state, the mutual
information available to legitimate parties decreases as
the time offset between two reciprocal channel measurements
increases.
Therefore, the secret key generation rate can be regarded as a function of the time lag $\Delta\ge 0$. When $\Delta=0$, the model degenerates into the ideal synchronous channel reciprocity scenario.

When $\xi^t =s$, node Eve is in sleeping mode, according to~\cite{9663196}, the secret key generation rate $R_{k}^{s}$ can be given by
\begin{equation}
R_{k}^{s}
=\log_{2}{\frac{\mathrm{det}(\boldsymbol{\Omega}_{a}^0)\mathrm{det}(\boldsymbol{\Omega}_{b}^0)}{\mathrm{det} (\boldsymbol{\Omega}_{ab}^{\Delta})}},
\label{6}
\end{equation}
where the covariance matrices are defined as
$\boldsymbol{\Omega}_{p_1...p_M}^{\Delta}=\mathbb{E}\{\boldsymbol{q}^{t}(\boldsymbol{q}^{t-\Delta})^H\}$, $\boldsymbol{q}=[p_1,p_2,...,p_M]^T$. 
Although instantaneous channel reciprocity is impaired by duplexing delay and CSI estimation errors, the underlying wireless channel remains reciprocal in a statistical sense, which suffices for PLKG. Thus, we can rewrite~\eqref{6} as
\begin{equation}
R_{k}^{s}
=2\log_{2}{\mathrm{det}(\boldsymbol{\Omega}_{a}^0)}-\log_{2}{\mathrm{det}(\boldsymbol{\Omega}_{ab}^{\Delta})},
\end{equation}
where the determinants of the covariance matrices are
\begin{align}
\mathrm{det}(\boldsymbol{\Omega}_{a}^0) &= \Omega _{a}^0 + \sigma^{2},  \nonumber \\
\mathrm{det}(\boldsymbol{\Omega}_{ab}^{\Delta}) &= (\Omega _{a}^0+\sigma^{2})^2 - |\Omega^{\Delta}_{a}|^2,
\label{7}
\end{align}
where $\Omega_{a}^0=\mathbb{E}\{\tilde{h}_{ab}^t(\tilde{h}_{ab}^t)^*\}=P_a\mathbb{E}\{|\boldsymbol{w}_{b}^H\boldsymbol{H}_{ab}\boldsymbol{w}_{a}|^2\}$ is the lag-$0$ legitimate channel autocorrelation, 
$\Omega^{\Delta}_{a}=\mathbb{E}\{\tilde{h}_{ab}^t(\tilde{h}_{ab}^{t-\Delta})^*\}=\rho^\Delta\Omega_{a}^0$ is the lag-$\Delta$ legitimate channel autocorrelation, 
the total noise variance is combined into $\sigma^{2}=\sigma^{2}_z+ \sigma^{2}_\zeta$. 
For the AR(1) model, the lag-$\Delta$ autocorrelation can be proven from the recursive relation as
\begin{align}
\Omega^{\Delta}_{a}&=\mathbb{E}\{\tilde{h}_{ab}^t(\tilde{h}_{ab}^{t-\Delta})^*\}=\mathbb{E}\{(\rho\tilde{h}_{ab}^{t-1}+\tilde{z}_{ab}^t)(\tilde{h}_{ab}^{t-\Delta})^*\}\nonumber\\
&=\rho\mathbb{E}\{\tilde{h}_{ab}^{t-1}(\tilde{h}_{ab}^{t-\Delta})^*\}=\rho\Omega^{\Delta-1}_{a}.
\end{align}
By further substituting the expression of $R_{k}^{s}$, we obtain ~\eqref{Rks}, which is provided at the top of the next page. 

\begin{figure*}[t]  
\begin{flalign}
R_k^s&=2\log_2{\big(P_a\mathbb{E}\{|\boldsymbol{w}_{b}^H\boldsymbol{H}_{ab}\boldsymbol{w}_{a}|^2\}+\sigma^{2}\big)
-\log_2\Big(\big(P_a\mathbb{E}\{|\boldsymbol{w}_{b}^H\boldsymbol{H}_{ab}\boldsymbol{w}_{a}|^2\}+\sigma^{2}\big)^2-\big|\rho^\Delta P_a\mathbb{E}\{|\boldsymbol{w}_{b}^H\boldsymbol{H}_{ab}\boldsymbol{w}_{a}|^2\}\big|^2}\Big).&
\label{Rks}
\end{flalign}
\vspace{-30pt}  
\end{figure*}

When $\xi^t =e$, node Eve is in eavesdropping mode, the secret key generation rate must be determined by evaluating the conditional mutual information $\mathcal{I}(\tilde{h}_{ab};\tilde{h}_{ba}|\tilde{h}_{ae}) $. It can be specifically expressed as
\begin{equation}
R_{k}^{e}
= \log_{2}{\frac{\mathrm{det}(\boldsymbol{\Omega}_{ae}^0)\mathrm{det}(\boldsymbol{\Omega}_{be}^0)}{\mathrm{det}(\boldsymbol{\Omega}_{e}^0)\mathrm{det}(\boldsymbol{\Omega}_{abe}^{\Delta})}}.
\label{Rk1}
\end{equation}
Similarly, due to the reciprocity of the channel, we can rewrite~\eqref{Rk1} as
\begin{equation}
R_{k}^{e}
=2\log_{2}{\mathrm{det}(\boldsymbol{\Omega}_{ae}^0)}-\log_{2}{\mathrm{det}(\boldsymbol{\Omega}_{e}^0)}-\log_{2}{\mathrm{det}(\boldsymbol{\Omega}_{abe}^{\Delta})},
\end{equation}
where the determinants of the covariance matrices are
\begin{align}
    \mathrm{det}(\boldsymbol{\Omega}_{e}^0) &= \Omega _{e}^0+\sigma^{2}_z,   \nonumber\\    
    \mathrm{det}(\boldsymbol{\Omega}_{ae}^0) &= \left(\Omega_{a}^0+\sigma^{2}\right)\left(\Omega_{e}^0+\sigma_{z}^{2}\right)-\left|\Omega_{ae}^0\right|^{2},  \\
    \mathrm{det}(\boldsymbol{\Omega}_{abe}^{\Delta}) &= 
   \big((\Omega_{a}^0+\sigma^{2})^{2}-|\Omega_{a}^{\Delta}|^{2} \big)\left(\Omega_{e}^0+\sigma_{z}^{2}\right)-2\sigma^{2}|\Omega_{ae}^0|^{2}, \nonumber
\end{align}
where $\Omega _{e}^0=\mathbb{E}\{\tilde{h}_{ae}^t(\tilde{h}_{ae}^t)^*\}=P_a\mathbb{E}\{|\boldsymbol{h}_{ae}\boldsymbol{w}_a|^2\}$ represents the lag-$0$ eavesdropping channel autocorrelation,  $\Omega _{ae}^0 =\mathbb{E}\{\tilde{h}_{ab}^t(\tilde{h}_{ae}^t)^*\}
=P_a\mathbb{E}\{\boldsymbol{w}_{b}^H\boldsymbol{H}_{ab}\boldsymbol{w}_a\boldsymbol{w}_a^H\boldsymbol{h}_{ae}^H\}$ is the lag-$0$ cross-correlation between legitimate and eavesdropping channels. 
In the analysis of information leakage, we do not explicitly model the time offset of the Eve's observations, because  Eve has no protocol or time constraints. Thus, we assume that Eve can obtain channel observations at the most advantageous time points, which sets an upper bound on the amount of information an eavesdropper can acquire, thereby preventing overestimation of the system's security performance. 
\begin{figure*}[t]  
\begin{flalign}
&R_{k}^{e}\!=\!2\log_{2}\!\Big(\big(P_a\mathbb{E}\{|\boldsymbol{w}_{b}^H\!\boldsymbol{H}_{ab}\boldsymbol{w}_{a}|^2\}\!+\!\sigma^{2}\big)\!\big(P_a\mathbb{E}\{|\boldsymbol{h}_{ae}\boldsymbol{w}_a|^2\}\!+\!\sigma_{z}^{2}\big)\!-\!\big|P_a\mathbb{E}\{\boldsymbol{w}_{b}^H\!\boldsymbol{H}_{ab}\boldsymbol{w}_a\boldsymbol{w}_a^H\boldsymbol{h}_{ae}^H\}\big|^{2}\Big)\!\!-\!\log_{2}\!\big(P_a\mathbb{E}\{|\boldsymbol{h}_{ae}\boldsymbol{w}_a|^2\!\!+\!\sigma_{z}^{2}\big)
 &\nonumber\\&\!-\!\log_{2}\!\Big(\big(\big(P_a\mathbb{E}\{|\boldsymbol{w}_{b}^H\!\boldsymbol{H}_{ab}\boldsymbol{w}_{a}|^2\}\!+\!\sigma^{2}\big)^2\!\!\!\!-\!\big|\rho^\Delta \!P_a\mathbb{E}\{|\boldsymbol{w}_{b}^H\!\boldsymbol{H}_{ab}\boldsymbol{w}_{a}|^2\}\big|^2\big)\!\big(P_a\mathbb{E}\{|\boldsymbol{h}_{ae}\boldsymbol{w}_a|^2\!\!+\!\sigma_{z}^{2}\big)\!\!-\!2\sigma^{2}\big|P_a\mathbb{E}\{\boldsymbol{w}_{b}^H\!\boldsymbol{H}_{ab}\boldsymbol{w}_a\boldsymbol{w}_a^H\boldsymbol{h}_{ae}^H\}\big|^{2}\Big).&
\label{Rke}
\end{flalign}
\vspace{-5pt}
\hrule  
\vspace{-10pt}  
\end{figure*}
Therefore, $R_{k}^{e}$ can be formulated as in~\eqref{Rke}, which is also provided at the top of the next page. 


\subsection{Data Transmission Rate}
Based on the Shannon-Hartley theorem~\cite{1697831}, the theoretical upper limit of the data transmission rate $R_{d}$ is determined by the channel capacity $C=B\log_{2}{\left ( 1+\mathrm{SNR}\right )}$, where $B$ is the channel bandwidth and $\mathrm{SNR}$ is the signal-to-noise ratio (SNR). Therefore, the data transmission rate $R_{d}$ at Alice is given by
\begin{equation}
R_{d}
=B\log_{2}{\left ( 1+\frac{P_a\mathbb{E}\{|\boldsymbol{w}_{b}^{H}\boldsymbol{H}_{ab}\boldsymbol{w}_{a}|^2\}}{\sigma^{2}}\right )}.
\end{equation}

\subsection{Optimization Problem}
Our objective is to jointly maximize the secret key generation rate $R_{k}$ and the data transmission rate $R_{d}$ under the potential threat of a smart eavesdropper. 
Thus, we optimize the beamforming vectors $\boldsymbol{w}_{a}$ and $\boldsymbol{w}_{b}$ by designing their amplitude coefficients $u_{n}$, $v_{n}$ and phase shift $\psi_n$, $\phi_n$, thereby enhancing the dual performance of both data transmission and secret key generation.
The optimization problem can be written as
\begin{subequations}
\begin{align}
\mathbb{P}:\quad\max_{\boldsymbol{w}_{a},\boldsymbol{w}_{b}} &\quad \lambda_{k}{R}_{k}^{\xi}+(1-\lambda_{k}){R}_{d},  \\
\text{s.t.} \quad
&\quad\|\boldsymbol{w}_{a}\|^2 =\|\boldsymbol{w}_{b}\|^2=1, \label{wa} \\
&\quad\;P_a, P_b \in [0, P_{\max}], \label{P} \\
&\quad\;\xi^t \in \{e,s\}, \label{xi}
\end{align}
\end{subequations}
where $\lambda_{k}\in [0,1]$ is the weighting factor that balances the trade-off between data transmission and secret
key generation.
Constraint~\eqref{wa} enforces the unit-norm beamforming design for Alice and Bob, constraint~\eqref{P} limits the transmit power within the maximum allowable budget $P_{\max}$, and constraint \eqref{xi} captures the two-mode behavioral model of the intelligent Eve.
Given the dynamic and temporally-coupled nature of this optimization problem, traditional convex optimization methods are ineffective. Therefore, we adopt a DRL approach to address this challenge.

\section{Problem solutions}
To solve the joint optimization problem of PLKG and data transmission under eavesdropping attacks, we propose a multi-agent SAC framework with LSTM adversary prediction. 

For the legitimate user Alice, each decision cycle begins with observing the current attacker mode $\xi^t$ and CSI $\tilde{h}_{ab}^t$, $\tilde{h}_{ae}^t$ determined by the preceding beamforming vectors $\boldsymbol{w}_{a}^{t-1}$ and $\boldsymbol{w}_{b}^{t-1}$.
Then, Alice and Bob choose their current beamforming vectors $\boldsymbol{w}_{a}^t$ and $\boldsymbol{w}_{b}^t$ as feedback to the environmental observation. This selection, in turn, causes the CSI to transition to $\tilde{h}_{ab}^{t+1}$ and $\tilde{h}_{ae}^{t+1}$. The described dynamic, where each action alters the subsequent state, clearly manifests a Markov decision process (MDP) governing Alice's interactive policy.

Since DRL is an effective method for solving MDP compliant time series optimization problems, we adopt the SAC algorithm as the DRL policy to optimize beamforming design. Consider Alice and Bob as multi-agent systems that perceive the dynamic environment around them in each time step and adjust beamforming vectors based on environmental feedback.

\subsection{MDP Formulation}
We define the MDP as a tuple $(\mathcal{S},\mathcal{A},r)$, where $\mathcal{S}$ denotes the state space, $\mathcal{A}$ denotes the action space, and $r$ represents the reward function. The state, action and reward are defined as follows:
\begin{enumerate}[label=(\arabic*)]
    \item  The environmental state $\mathbf{s}^t \in \mathcal{S}$ represents the wireless channel conditions among Alice, Bob, and Eve. At time step $t$, Alice observes the current CSI and attacker mode:
    \begin{equation}
    \mathbf{s}^t=\{\tilde{{h}}_{ab}^t,\tilde{{h}}_{ae}^t,\xi^t\}.
    \end{equation}
    \item The joint action $\mathbf{a}^t \in \mathcal{A}$ of Alice and Bob at time step $t$ is defined as their beamforming vectors:
    \begin{equation}
    \mathbf{a}^t=\{\mathbf{a}_{a}^t,\mathbf{a}_{b}^t\}=
\{\boldsymbol{w}_{a}^t,\boldsymbol{w}_{b}^t\}. 
    \end{equation}
    \item The instant reward $r$ is designed to simultaneously consider the secret key generation rate $R_k$ and the data transmission rate $R_{d}$:
    \begin{equation}
r^t=\lambda_{k}({R}_{k}^\xi)^{t}+(1-\lambda_{k}){R}_{d}^t.
\end{equation}
\end{enumerate}

\subsection{Multi-Agent SAC}
In this work, we adopt the SAC algorithm as the core DRL framework to optimize the beamforming vectors in a multi-agent setting. SAC is a model-free, off-policy actor-critic algorithm that combines the benefits of maximum entropy DRL with stable policy optimization. 
The objective function in SAC is defined as
\begin{equation}
J(\pi)=\sum_{t=0}^T\mathbb{E}_{\mathbf{s}^t,\mathbf{a}^t}\left[r(\mathbf{s}^t,\mathbf{a}^t)+\alpha\mathcal{H}\big(\pi(\cdot|\mathbf{s}^t)\big)\right],
\end{equation}
where $\alpha$ is the temperature parameter that controls the trade-off between exploration and exploitation, and $\mathcal{H}\big(\pi(\cdot|\mathbf{s}^t)\big)$ is the entropy of the policy in state $\mathbf{s}^t$. The maximum entropy objective stabilizes convergence in non-stationary AR(1) channels by preventing premature policy determinism. 

We model Alice and Bob as two cooperative agents that jointly optimize their beamforming vectors $\boldsymbol{w}_{a}$ and $\boldsymbol{w}_{b}$. Each agent uses its own actor network $\pi_{\theta}(\mathbf{s})$ to generate actions and shares two common critic networks $Q_{\omega_1}(\mathbf{s},\mathbf{a})$ and $Q_{\omega_2}(\mathbf{s},\mathbf{a})$ to evaluate the joint action.
The loss function of any $Q$ function is given in~\eqref{LQ} at the top of the next page, 
\begin{figure*}[t]  
\begin{equation}
L_{Q}(\omega) = \mathbb{E}_{\mathbf{s}^{t},\mathbf{a}^{t},r^{t},\mathbf{s}^{t+1},\mathbf{a}^{t+1}}
\bigg[ \frac{1}{2}\Big( Q_{\omega}(\mathbf{s}^{t},\mathbf{a}^{t}) - \Big( r^{t} + \gamma \big( \min_{j=1,2} Q_{\omega_{j}^{-}} (\mathbf{s}^{t+1},\mathbf{a}^{t+1}) - \alpha \log \pi (\mathbf{a}^{t+1}|\mathbf{s}^{t+1}) \big) \Big) \Big)^{2} \bigg].
\label{LQ}
\end{equation}
\hrule  
\vspace{-10pt}  
\end{figure*}
where $\gamma$ is the discount factor,
$(\mathbf{s}^{t},\mathbf{a}^{t},r^{t},\mathbf{s}^{t+1})$ are sampled from the data $\mathcal{D}$ collected by the policy in the past, $\mathbf{a}^{t+1}$ is sampled from the policy $\pi_\theta (\cdot|\mathbf{s}^{t+1})$, and two target Q-networks $Q_{\omega^-}$ correspond to two Q-networks to enhance training stability.

Meanwhile, the loss function for policy $\pi$ can be derived from the Kullback-Leibler divergence, simplified to
\begin{equation}
L_{\pi}\left(\theta\right)=\mathbb{E}_{\mathbf{s}^{t},\mathbf{a}^{t}}\left[\alpha\log\left(\pi_{\theta}\left(\mathbf{a}^{t}|\mathbf{s}^{t}\right)\right)-\min_{j=1,2}Q_{\omega_j}\left(\mathbf{s}^{t},\mathbf{a}^{t}\right)\right],
\end{equation}
where 
$\mathbf{s}^{t}$ and $\mathbf{a}^{t}$ sampled from $\mathcal{D}$ and $\pi_\theta$ respectively.

In addition, to make the entropy regularization coefficient in the SAC algorithm adaptive, we derive the loss function for $\alpha$ by ensuring the mean entropy exceeds $\mathcal{H}_0$:
\begin{equation}
L(\alpha)=\mathbb{E}_{\mathbf{s}^t,\mathbf{a}^t}\left[-\alpha\log\pi_\theta(\mathbf{a}^t|\mathbf{s}^t)-\alpha\mathcal{H}_0\right].
\end{equation}
When the policy entropy falls below the target value $\mathcal{H}_0$, the importance of increasing policy entropy is elevated; conversely, when the policy entropy rises above $\mathcal{H}_0$, the focus shifts toward value enhancement.

\subsection{LSTM Adversary Prediction}
Another crucial problem is that legitimate users cannot observe the CSI of the eavesdropper. 
To address this partial observability issue, we introduce a friendly node called Fred deployed near Eve. This node Fred, capable of communicating legitimately with both Alice and Bob, provides channel measurements that are highly correlated with Eve's channel due to their spatial proximity. By leveraging Fred's historical channel data, we can more accurately predict both the eavesdropping channel $\tilde{{h}}_{ae}$ and the attacker mode $\xi$, enhancing robustness and performance in partially observable environments.

Specifically, we deploy a single-antenna node Fred in the system. The channels from Fred to Alice and Bob, denoted as $\tilde{h}_{af}^t$ and $\tilde{h}_{bf}^t$, serve as input to an LSTM network: 

\begin{equation}
\mathbf{x}^t=[\tilde{{h}}_{af}^t,\tilde{{h}}_{bf}^t].
\end{equation}
The LSTM processes this sequence and outputs a joint prediction containing both the complex-valued eavesdropping channel and the attacker mode state, formulated as
\begin{equation}
[\hat{h}_{ae}^t,\hat{\xi}^t]=\mathrm{LSTM}(\mathbf{x}^{t-L},...,\mathbf{x}^t),
\end{equation}
where $L$ is the sequence length. These predicted values are then used to form the comprehensive partial observation state for the SAC agent, which is denoted as
\begin{equation}
\hat{\mathbf{s}}^t=\{\tilde{{h}}_{ab,i}^t,\hat{{h}}_{ae}^t,\hat\xi^t\}.
\end{equation}



\section{Simulation Results and Discussion}
We conduct numerical simulations to evaluate the performance of the proposed multi-agent SAC algorithm with LSTM-based adversary prediction. The simulation environment is implemented in Python with PyTorch 2.7.0. 
The legitimate transceivers Alice and Bob are each equipped with $N=8$ antennas, and the beamforming power is constrained to $P_a=P_b=P=20\ \mathrm{dBm}$.
The action dimension for each agent is thus $2N=16$, representing the real and imaginary components of their respective beamforming vectors $\boldsymbol{w}_{a}$ and $\boldsymbol{w}_{b}$. 
The state dimension is set to $5$, including the real and imaginary parts of the legitimate channel $\tilde{{h}}_{ab}^t$, eavesdropping channel $\hat{{h}}_{ae}^t$, and  the mode of the attacker  $\hat{\xi}^t$.


The SAC agent is built upon a fully-connected neural network architecture for both the actor and critic networks. Specifically, the actor and each critic network consist of two hidden layers, each comprising $512$ neurons, with Layer Normalization applied after each linear layer to stabilize training. Simultaneously, the LSTM-based adversary prediction module is designed with a single LSTM layer of hidden size $64$, followed by a fully-connected output layer. The discount factor $\gamma$ is set to $0.99$, target network update rate to $0.005$, and learning rates for both actor and critic to $10^{-4}$. The entropy temperature $\alpha$ is initialized at $0.02$.


\begin{figure}[t]
    \centering 
    \includegraphics[width=0.35\textwidth]{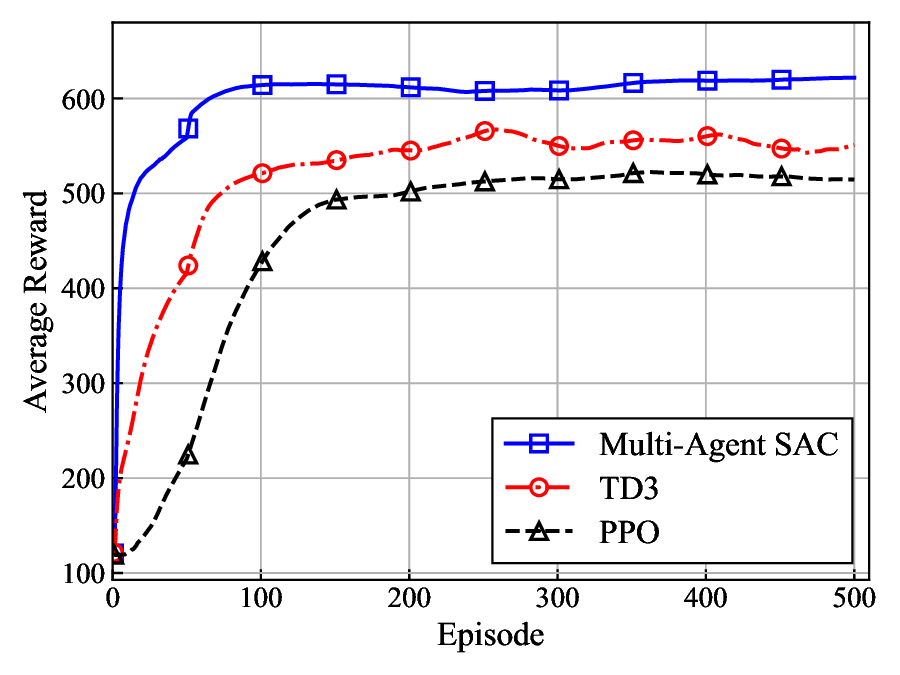} 
     \vspace{-10pt} 
    \caption{Comparison of rewards between the proposed method, TD3 and PPO.} 
    \label{fig2} 
        \vspace{-10pt}
\end{figure}
\begin{figure}[t]
    \centering 
    \includegraphics[width=0.35\textwidth]{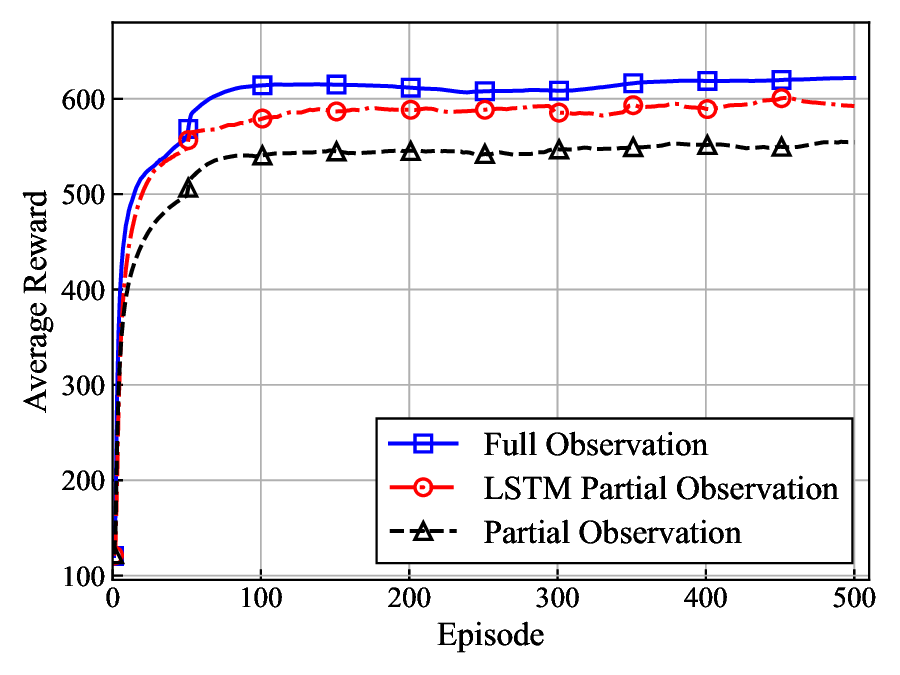} 
     \vspace{-10pt} 
    \caption{The reward for full observation and partial observation information of Multi-Agent SAC algorithm.} 
    \label{fig3} 
        \vspace{-15pt}
\end{figure}
We first compare the proposed SAC algorithm with other actor-critic based DRL algorithms, including Twin Delayed Deep Deterministic Policy Gradient (TD3) and Proximal Policy Optimization (PPO), in terms of average training reward. 
As illustrated in~\cref{fig2}, SAC achieves fast convergence and high stability, yielding an average reward of approximately $615$. This performance gain stems from its entropy-regularized exploration policy, which effectively balances exploration and exploitation in complex environments. The minor fluctuations in reward observed after convergence are mainly due to environmental stochasticity, entropy regularization, and inherent variability in the reward function. In comparison, both PPO and TD3 exhibit inferior performance relative to SAC, underscoring the latter's advantages in tasks such as beamforming optimization, secret key generation, and data transmission.


In \cref{fig3}, we present a comparison between full and partial observation scenarios, demonstrating the effectiveness of LSTM-based prediction. Under partial observation, the reward loss is controlled within $4\%$, significantly outperforming the original SAC algorithm. In contrast, due to the lack of mode information regarding Eve and the eavesdropping channel, the original SAC algorithm exhibits substantial errors, with an average performance degradation of approximately $15\%$. This high-precision prediction boosts beamforming to reduce leakage, steadily raising key and data rates.
\begin{figure}[t]
    \centering 
    \includegraphics[width=0.32\textwidth]{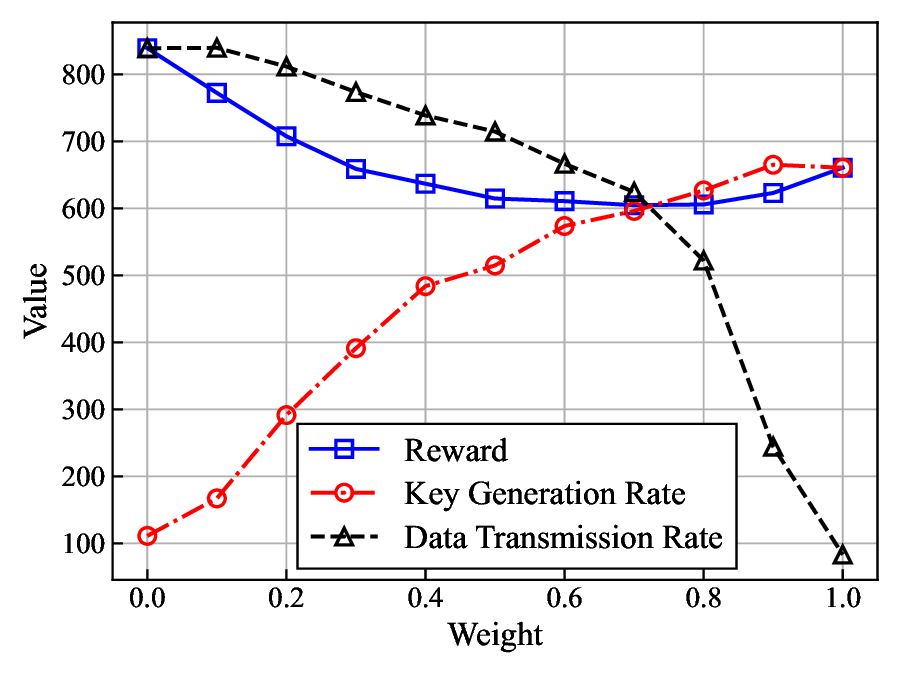} 
     \vspace{-10pt} 
    \caption{The reward, the key generation rate and the data transmission rate with different weight factors $\lambda_{k}$.} 
    \label{fig4} 
        \vspace{-10pt}
\end{figure}
\begin{figure}[t]
    \centering 
    \includegraphics[width=0.32\textwidth]{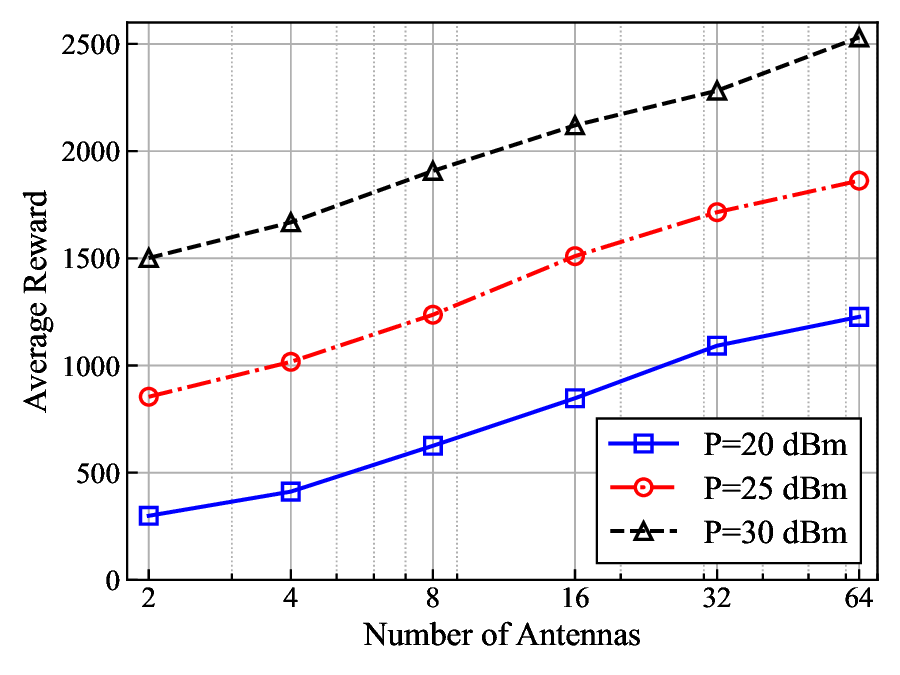} 
     \vspace{-10pt} 
    \caption{The reward versus the number of antennas $N$ under different power constraints $P$.} 
    \label{fig5} 
        \vspace{-15pt}
\end{figure}

As shown in~\cref{fig4}, different weight factor $\lambda_k$ settings correspond to distinct design objectives in the dual-goal scenario of simultaneous secret key generation and data transmission. Specifically, setting $\lambda_k=0$ focuses exclusively on maximizing the data transmission rate, whereas $\lambda_k=1$ prioritizes the maximization of the secret key generation rate. As $\lambda_k$ varies continuously from $0$ to $1$, the overall reward exhibits a non-monotonic behavior, first decreasing and then increasing. 
Notably, when $\lambda_k$ is adjusted from $1$ to $0.9$, both the data transmission rate and the secret key generation rate are improved simultaneously, indicating a mutually beneficial effect between the two objectives.

~\cref{fig5} shows the average reward obtained as the number of antennas $N$ varies for different transmission power constraints $P$. The reward increases with more antennas at all power levels, confirming that larger arrays provide higher beamforming gain and spatial freedom, thereby improving both data transmission rate and secret key generation rate. Meanwhile, for a fixed antenna size, reward values rise with transmission power, consistent with the fact that higher power enhances SNR and strengthens equivalent channel characteristics. 

\section{Conclusion}
We have investigated the joint optimization of beamforming vectors for simultaneous secret key generation and data transmission in a multi-antenna TDD system under intelligent eavesdropping attacks. By modeling the legitimate channel as a temporally correlated AR(1) process, 
the resulting joint optimization problem was shown to be highly nonconvex and temporally coupled. A multi-agent SAC framework was developed to address this dynamic and non-convex problem. To tackle the issue of partial observability, an LSTM-based prediction module was integrated into the framework.  Numerical simulations confirmed the superiority of the proposed approach over other DRL algorithms, achieving higher average rewards with faster convergence and greater stability. This work provides a robust and efficient solution for securing future 6G communication systems through the intelligent co-design of secret key generation and data transmission.

\bibliographystyle{ieeetr}

\balance
\end{document}